\makeatletter
\input{filecontents.sty}
\makeatother

\documentclass[%
secnumarabic,
showpacs,
showkeys,
twoside,
floatfix,
a4paper,
superscriptaddress,
groupedaddress,
amsmath,amssymb,amsfonts,amsbsy,
aps,reprint,onecolumn,notitlepage]{preprint}

\makeatletter
\def\lrbox#1{%
  \edef\reserved@a{%
    \endgroup
    \global\setbox#1\hbox{%
      \begingroup\aftergroup}%
    \def\noexpand\@currenvir{\@currenvir}%
    \def\noexpand\@currenvline{\on@line}}%
  \reserved@a
  \@endpefalse
  \color@setgroup
  \ignorespaces}
\def\endlrbox{\unskip\color@endgroup}
\makeatother

\renewcommand\documentclass[2][]{}
\newcommand{\changefont}[3]{}
\newcommand{\journal}[1]{}
\newbox\mybox
\newenvironment{frontmatter}{}{\maketitle}
\def\sep{\unskip, }%
\let\ead=\email
\let\PACS=\pacs
\newenvironment{keyword}{\begin{lrbox}{\mybox}\footnotesize}{\end{lrbox}\keywords{\usebox{\mybox}}}
\AtBeginDocument{%
  \IfFileExists{date.tex}{%
    \input{date.tex}

  }
}
\newcommand{\secreferences}{}
\documentclass[fleqn,final,3p,times]{elsarticle}

\usepackage[T1]{fontenc}
\usepackage{graphicx}
\graphicspath{{figures/}}
\usepackage{amsmath}
\usepackage{braket}
\usepackage{tabularx}
\usepackage[pdftitle={A stability analysis of a real space split operator method for the Klein-Gordon equation},
pdfauthor={Frederick Blumenthal and Heiko Bauke},
pdfcreator={},
pdfsubject={},
pdfproducer={},
pdfkeywords={Klein-Gordon equation; split-operator-method; stability analysis},
hyperfootnotes=false,naturalnames]{hyperref}

\journal{Journal of Computational Physics}

\renewcommand*{\vec}[1]{\boldsymbol{#1}}

\newcommand*{\op}[1]{{\hat#1}}
\newcommand*{\I}{\mathrm{i}}
\newcommand*{\D}{\mathrm{d}}
\newcommand*{\e}{\mathrm{e}}
\newcommand*{\bigO}[1]{O\left(#1\right)}
\newcommand*{\norm}[1]{\left\lVert#1\right\rVert}
\newcommand*{\abs}[1]{\left\lvert#1\right\rvert}

\providecommand{\secreferences}{\section*{References}}

\begin{document}

\begin{frontmatter}
  \title{A stability analysis of a real space split operator method for~the~Klein-Gordon~equation}
  \author{Frederick Blumenthal}
  \author{Heiko Bauke}
  \ead{heiko.bauke@mpi-hd.mpg.de}
  \address{Max-Planck-Institut f{\"u}r Kernphysik, Saupfercheckweg~1,
    69117~Heidelberg, Germany}
  \begin{abstract}
    We carry out a stability analysis for the real space split operator
    method for the propagation of the time-dependent Klein-Gordon
    equation that has been proposed Ruf et al. [M.~Ruf, H.~Bauke,
    C.H.~Keitel, A real space split operator method for the Klein-Gordon
    equation, Journal of Computational Physics 228 (24) (2009)
    9092--9106,
    \href{http://dx.doi.org/10.1016/j.jcp.2009.09.012}{doi:10.1016/j.jcp.2009.09.012}]. The
    region of algebraic stability is determined analytically by means of
    a von-Neumann stability analysis for systems with homogeneous scalar
    and vector potentials.  Algebraic stability implies convergence of
    the real space split operator method for smooth absolutely
    integrable initial conditions.  In the limit of small spatial grid
    spacings $h$ in each of the $d$ spatial dimensions and small
    temporal steps $\tau$, the stability condition becomes
    $h/\tau>\sqrt{d}c$ for second order finite differences and
    $\sqrt{3}h/(2\tau)>\sqrt{d}c$ for fourth order finite differences,
    respectively, with $c$ denoting the speed of light.  Furthermore, we
    demonstrate numerically that the stability region for systems with
    inhomogeneous potentials coincides almost with the region of
    algebraic stability for homogeneous potentials.
  \end{abstract}
\end{frontmatter}

\section{Introduction}
\label{sec:intro}

The Klein-Gordon equation is a partial differential equation that
governs the quantum evolution of wave functions for relativistic
spinless particles. It finds its applications for example in modeling
light matter interaction in the relativistic regime
\cite{Protopapas:etal:1997:Atomic_physics,Salamin:etal:2006:Relativistic_laser-matter,Brabec:2009:Strong_Field_Laser_Physics}.
The derivation of analytical solutions for the time-dependent problems
in relativistic quantum dynamics requires sophisticated mathematical
tools
\cite{Bagrov:Gitman:1990:Exact_Solutions,Gross:1999:Relativistic_QM,Strange:1998:Relativistic_QM,Schwabl:2008:Advanced_QM,Wachter:2011:Relativistic_QM,Ehlotzky:etal:2009:Fundamental_processes}
and many setups necessitate numerical approaches solving the Dirac
equation
\cite{Kylstra:etal:1997:1D_Dirac,Rathe:elal:1997:Intense_laser-atom,Braun:etal:1999:Numerical_approach,Mocken:Keitel:2004:Quantum_dynamics,Mocken:Keitel:2008:FFT-split-operator,Selsto:etal:2009:Solution_of_the_Dirac_equation,Bauke:etal:2011:Relativistic_ionization,Bauke:Keitel:2011:Accelerating}
or the Klein-Gordon equation
\cite{Bauke:etal:2011:Relativistic_ionization,Taieb:etal:1998:Signature},
especially problems with time-dependent potentials. In weakly
relativistic regimes the solution of the Pauli equation with first order
relativistic corrections may be sufficient
\cite{Hu:Keitel:2001:ions_in_intense_laser_fields}.

Ruf et~al.~\cite{Ruf:2009:Klein-Gordon_Splitoperator} developed an
explicit real space split operator finite difference scheme for the
numerical solution of the time-dependent Klein-Gordon equation. It has
been applied for example to study relativistic ionization
characteristics of laser-driven hydrogen like ions
\cite{Bauke:etal:2011:Relativistic_ionization} and numerical signatures
of non-self-adjointness in quantum Hamiltonians
\cite{Ruf:etal:2011:Numerical_signatures}.  As an explicit scheme, the
real space split operator finite difference scheme allows for an
efficient implementation on distributed memory parallel
computers. Explicit finite difference schemes for hyperbolic partial
differential equations, however, are known to be never unconditionally
stable \cite{Courant:Friedrichs:Lewy:1928:PDE,Strikwerda:2004:FDS}. This
means, that the numerical error that is caused by the application of an
explicit finite difference scheme to the Klein-Gordon equation will grow
exponentially unless the spatial and temporal grid spacings fulfill
certain conditions. The purpose of this contribution is to study the
stability of the explicit real space split operator finite difference
scheme as introduced in \cite{Ruf:2009:Klein-Gordon_Splitoperator} and
to derive stability conditions for this scheme.

The reminder of the paper is organized as follows. In order to keep the
paper self-contained, we will give a brief description of the real space
split operator method for the Klein-Gordon equation in
Sec.~\ref{sec:split}. Section~\ref{sec:stability} will state more
precisely the notion of numerical stability by defining strong and
algebraic stability, while Sec.~\ref{sec:Conditions} reviews conditions
for these notions of stability.  A von-Neumann stability analysis for
constant potentials will be performed in Sec.~\ref{sec:Neumann} and in
Sec.~\ref{sec:numerics} we will compare our theoretical results with the
stability of an actual computer implementation of the real space split
operator method for systems with homogeneous potentials as well as for
systems with spatial dependent potentials.

\section{A real space split operator method for the Klein-Gordon equation}
\label{sec:split}

The Klein-Gordon equation---sometimes also called Klein-Fock-Gordon or
Klein-Gordon-Schr{\"o}dinger equation---is an equation of motion for a
complex-valued scalar quantum wave function $\varphi(\vec x, t)$
\cite{Wachter:2011:Relativistic_QM,Feshbach:Villars:1954:Relativistic_Wave_Mechanics,Kragh:1984:Klein-Gordon}.
It governs the behavior of a charged spinless particle with mass $m$ and
charge $q$ moving under the effect of electrodynamic potentials $\vec
A(\vec x, t)$ and $\phi(\vec x, t)$.  Introducing the speed of light $c$
and the reduced Planck constant $\hbar$, the scalar Klein-Gordon
equation reads
\begin{equation}
  \label{eq:scalar_Klein-Gordon}
  \left[
    \left(
      \I\hbar\frac{\partial}{\partial t} - q\phi(\vec x, t)
    \right)^2 -
    c^2\left(
      -\I\hbar\nabla - q\vec A(\vec x, t)
    \right)^2 - m^2 c^4
  \right]\varphi(\vec x, t)=0\,.
\end{equation}
This hyperbolic partial differential equation is of second order in
time. For numerical purposes a transformation of
(\ref{eq:scalar_Klein-Gordon}) into a first order partial differential
equation is advantageous.  Introducing the two-component wave function
\cite{Feshbach:Villars:1954:Relativistic_Wave_Mechanics}
\begin{equation}
  \Psi(\vec x, t) =
  \begin{pmatrix}
    \Psi_1(\vec x, t) \\ \Psi_2(\vec x, t)
  \end{pmatrix} =
  \begin{pmatrix}
    \tfrac12\left(
      \varphi(\vec x, t) + \tfrac{1}{mc^2} \left(
        \I\hbar\tfrac{\partial}{\partial t}
        - q\phi(\vec x, t)
      \right)\varphi(\vec x, t)
    \right) \\[2ex]
    \tfrac12\left(
      \varphi(\vec x, t) - \tfrac{1}{mc^2} \left(
        \I\hbar\tfrac{\partial}{\partial t}
        - q\phi(\vec x, t)
      \right)\varphi(\vec x, t)
    \right)
  \end{pmatrix}
\end{equation}
the Klein-Gordon equation (\ref{eq:scalar_Klein-Gordon}) may be
transformed into Schr{\"o}dinger form with the Hamiltonian operator $\op H$,
viz.
\begin{equation}
  \label{eq:Klein-Gordon-Schroedinger}
  \I\hbar\frac{\partial \Psi(\vec x, t)}{\partial t}
   = \op H(t)\Psi(\vec x, t) =
  \left(
    \frac{\sigma_3+\I\sigma_2}{2m}
    \left( -\I\hbar\nabla - q \vec A(\vec x, t) \right)^2
    + q \phi(\vec x, t) + \sigma_3 mc^2
  \right)\Psi(\vec x, t)\,.
\end{equation}
The wave-function's components are related by the complex Hermitian
unitary Pauli matrices $\sigma_2$ and $\sigma_3$ that obey the Pauli algebra
\begin{subequations}
  \label{eq:pauli_algebra}
  \begin{align}
    \sigma_i \sigma_j & = \I\epsilon_{i,j,k} \sigma_k + \delta_{i,j} \,, \\
    [\sigma_i, \sigma_j] & = 2\I\epsilon_{i,j,k}\sigma_k\,, \\
    \{\sigma_i, \sigma_j\} & = 2\delta_{i,j}\,,
  \end{align}
\end{subequations}
with $i,j,k\in\{1,2,3\}$ and where $\epsilon_{i,j,k}$ denotes the
permutation symbol and $\delta_{i,j}$ is the Kronecker delta. The Pauli
algebra does not define the Pauli matrices uniquely. In numerical
applications, we choose the representation
\begin{equation}
  \label{eq:Pauli_matrices}
  \sigma_1 = \begin{pmatrix} 0&1\\ 1&0 \end{pmatrix}\,{,}\quad
  \sigma_2 = \begin{pmatrix} 0&-\I\\ \I&0 \end{pmatrix}\,{,}\quad
  \sigma_3 = \begin{pmatrix} 1&0\\ 0&-1 \end{pmatrix} \,.
\end{equation}
Solutions $\Psi(\vec x, t)$ of equation
(\ref{eq:Klein-Gordon-Schroedinger}) satisfy the continuity equation
\begin{equation}
  \frac{\partial\rho(\vec x, t)}{\partial t} + \nabla \cdot \vec j(\vec x, t) = 0
\end{equation}
with the density
\begin{equation}
  \label{eq:kg_dens}
  \rho(\vec x, t) = \Psi^*(\vec x, t)\sigma_3 \Psi(\vec x, t) =
  |\Psi_1(\vec x, t)|^2 - |\Psi_2(\vec x, t)|^2\,.
\end{equation}
and the current
\begin{equation}
  \label{eq:kg_curr}
  \vec j(\vec x, t) = - \frac{\I\hbar}{2m}\big[
    \Psi^*(\vec x, t) \sigma_3 (\sigma_3+\I\sigma_2)\nabla\Psi(\vec x, t) -
    (\nabla\Psi^*(\vec x, t)) \sigma_3 (\sigma_3+\I\sigma_2)\Psi(\vec x, t)
  \big] - 
  \frac{q\vec A(\vec x, t)}{m}\Psi^*(\vec x, t)\sigma_3 (\sigma_3+\I\sigma_2)\Psi(\vec x, t)\,,
\end{equation}
where $\Psi^*(\vec x, t)$ denotes the complex conjugate of $\Psi(\vec
x,t)$.  The density (\ref{eq:kg_dens}) is not positive definite and,
therefore, the integral 
\begin{equation}
  \label{eq:pseudo_norm}
  \norm{\Psi(\vec x, t)}_{\sigma_3}^2 =
  \int \Psi^*(\vec x, t)\sigma_3 \Psi(\vec x, t)\,\D x^d 
\end{equation}
over the \mbox{$d$-dimensional} space is not a norm and $\rho(\vec x,
t)$ cannot be interpreted as a probability density and $\vec j(\vec x,
t)$ is not a probability density current.  The quantity $q\rho(\vec x,
t)$, however, may be interpreted as a charge density and $q\,\vec j(\vec
x, t)$ as a charge current
\cite{Gross:1999:Relativistic_QM,Strange:1998:Relativistic_QM,Schwabl:2008:Advanced_QM,Wachter:2011:Relativistic_QM,Feshbach:Villars:1954:Relativistic_Wave_Mechanics}.
The conservation of the pseudo norm (\ref{eq:pseudo_norm}) implies
charge conservation.  Note that the Euclidean norm
\begin{equation}
  \label{eq:norm}
  \norm{\Psi(\vec x, t)}^2 =
  \int \Psi^*(\vec x, t) \Psi(\vec x, t)\,\D x^d 
\end{equation}
is not conserved under the evolution of the Klein-Gordon equation.

In order to derive the real space split operator method of
\cite{Ruf:2009:Klein-Gordon_Splitoperator}, we start from the formal
solution of (\ref{eq:Klein-Gordon-Schroedinger}) with the initial
condition $\Psi(\vec x, 0)$ that may be expressed as
\begin{equation}
  \Psi(\vec x, t)=\op U(t, 0) \Psi(\vec x, 0)
\end{equation}
by introducing the unitary time-evolution operator
\cite{Pechukas:Light:1966:Time-Displacement}
\begin{equation}
  \label{eq:U_op}
  \op U(t_2, t_1) =
  \op T\exp\left(
    -\frac{\I}{\hbar}\int_{t_1}^{t_2} \op H(t') \,\D t'
  \right)\,,
\end{equation}
and Dyson's time ordering operator $\hat T$. The time-evolution operator
$\op U(t_2, t_1)$ effects the wave-function's evolution from time $t_1$
to time~$t_2$. Let $\op O(t)$ denote some possibly time-dependent
operator and define the operator
\begin{equation}
  \label{eq:U_O_op}
  \op U_{\op O}(t_2, t_1, \delta) =
  \exp\left(
    -\delta\frac{\I}{\hbar}\int_{t_1}^{t_2} \op O(t') \,\D t'
  \right)\,,
\end{equation}
that depends on the times $t_1$ and $t_2$ and the auxiliary
parameter~$\delta$.  Expanding $ \op U\left(t + \tau , t\right)$ to
the third order in $\tau $ and introducing the operators
\begin{subequations}
  \begin{align}
    \op K(\vec x, t) & =
    \frac{\sigma_3+\I\sigma_2}{2m}
    \left( -\I\hbar\nabla - q \vec A(\vec x, t) \right)^2  \\
    \op V(\vec x, t) & =
    q \phi(\vec x, t) + \sigma_3 mc^2 \,,
  \end{align}
\end{subequations}
the time-evolution operator for the Klein-Gordon equation
(\ref{eq:U_op}) can be factorized as
\begin{align}
  \nonumber
  \op U\left(t + \tau , t\right) & = 
  \exp\left(
    -\frac{\I}{\hbar}\int_{t}^{t+\tau } \op H(t') \,\D t'
  \right) + O\left(\tau ^3\right) \\
  \label{eq:U_split}
  & =
  \op U_{\op V}\left(t + \tau , t, \tfrac12\right)
  \op U_{\op K}\left(t + \tau , t, 1\right)
  \op U_{\op V}\left(t + \tau , t, \tfrac12\right) + O\left(\tau ^3\right) \,.
\end{align}
Neglecting terms of order $O\left(\tau ^3\right)$, equation
(\ref{eq:U_split}) gives an explicit second order accurate time-stepping
scheme for the propagation of the Klein-Gordon wave function $ \Psi(\vec
x, t)$. 

The operator $\op U_{\op V}\left(t+\tau , t, \delta\right)$ is
diagonal in real space and explicitly given by 
\begin{equation}
  \label{eq:op_U_V}
  \op U_{\op V}\left(t+\tau , t, \delta\right) \Psi(\vec x, t) =
  \begin{pmatrix}
    \textstyle
    \exp\left(
      -\delta\frac{\I}{\hbar}\int_{t}^{t+\tau } q \phi(\vec x, t') + mc^2 \,\D t'
    \right)\Psi_1(\vec x, t) \\[2ex]
    \textstyle \exp\left(
      -\delta\frac{\I}{\hbar}\int_{t}^{t+\tau } q \phi(\vec x, t') - mc^2 \,\D t'
    \right)\Psi_2(\vec x, t)
  \end{pmatrix}\,.
\end{equation}
The action of the operator $\op U_{\op K}\left(t+\tau , t, \delta\right)$
on $\Psi(\vec x, t)$ may be calculated by the Taylor series of the
exponential function
\begin{align}
  \nonumber
  \op U_{\op K}\left(t+\tau , t, \delta\right) \Psi(\vec x, t)
  & =
  \exp\left(
    -\delta\frac{\I}{\hbar}\int_{t}^{t+\tau } \op K(\vec x, t') \,\D t'
  \right)\Psi(\vec x, t) \\[1ex]
  \label{eq:op_U_K_ser}
  & =
  \sum_{j=0}^\infty \frac{1}{j!}\left[
    -\delta\frac{\I}{\hbar}\int_{t}^{t+\tau } \frac{\sigma_3+\I\sigma_2}{2m}
    \left( \frac\hbar\I\nabla - q \vec A(\vec x, t') \right)^2 \,\D t'
  \right]^j \Psi(\vec x, t) \,.
  % & =
  % \sum_{j=0}^\infty \frac{(\sigma_3+\I\sigma_2)^j}{j!}\left[
  %   \frac{-\delta\I}{2m\hbar}\int_{t}^{t+\tau }
  %   \left( \frac\hbar\I\nabla - q \vec A(\vec x, t') \right)^2 \,\D t'
  % \right]^j \Psi(\vec x, t)\,. \\
  \intertext{Recognizing that all but the first two terms of the
    infinite series (\ref{eq:op_U_K_ser}) vanish because the operator
    $(\sigma_3+\I\sigma_2)$ is nilpotent, that is,
    \mbox{$(\sigma_3+\I\sigma_2)^2=0$}, yields}
  \label{eq:op_U_K_lin}
  \op U_{\op K}\left(t+\tau , t, \delta\right) \Psi(\vec x, t)
  & =
  \Psi(\vec x, t) - (\sigma_3+\I\sigma_2)
  \frac{\delta\I}{2m\hbar}\int_{t}^{t+\tau }
  \left( \frac\hbar\I\nabla - q \vec A(\vec x, t') \right)^2 \Psi(\vec x, t)\,\D t'
  \,.
\end{align}
Introducing the quantities
\begin{subequations}
  \label{eq:U_K_constants}
  \begin{align}
    \label{eq:U_K_constants_0}
    c_{0} & =
    \exp\left(
      -\frac{q\I}{2\hbar}\int_{t}^{t+\tau } \phi(\vec x, t') \,\D t'
    \right)\,, \\
    \label{eq:U_K_constants_1}
    c_{1} & =
    \frac{ \I \tau \hbar}{2m} \,, \\
    \label{eq:U_K_constants_2}
    \vec c_{2} & =
    \frac{q}{m} \int_{t}^{t+\tau } \vec A(\vec x, t') \,\D t' \,,\\
    \label{eq:U_K_constants_3}
    c_{3} & =
    \int_{t}^{t+\tau }
    \frac{ q}{2m}\nabla\cdot\vec A(\vec x, t') -
    \frac{\I q^2}{2m\hbar} \vec A(\vec x, t')^2
    \,\D t'\,,
  \end{align}
\end{subequations}
and taking advantage of the standard representation
(\ref{eq:Pauli_matrices}) of the Pauli matrices the operator product
(\ref{eq:U_split}) finally may be transformed into a product of the
compact matrices
\begin{multline}
  \label{eq:op_U_split2}
  \op U\left(t+\tau , t\right)
  % \begin{pmatrix}
  %   \Psi_1(\vec x, t) \\ \Psi_2(\vec x, t)
  % \end{pmatrix}
  =
  \overbrace{
    \begin{pmatrix}
      c_{0} & 0 \\
      0 & c_{0}
    \end{pmatrix}
  }^{(*)}
  %\overbrace{
    \begin{pmatrix}
      \e^{-\I mc^2\tau/(2\hbar)} & 0 \\
      0 & \e^{+\I mc^2\tau/(2\hbar)}
    \end{pmatrix}
  %}^{(**)}
  %\overbrace{
    \begin{pmatrix}
      1+(c_1 \nabla^2 + \vec c_2 \cdot \nabla + c_3) &
      \phantom{0}+(c_1 \nabla^2 + \vec c_2 \cdot \nabla + c_3) \\
      \phantom{0}-(c_1 \nabla^2 + \vec c_2 \cdot \nabla + c_3) &
      1-(c_1 \nabla^2 + \vec c_2 \cdot \nabla + c_3) 
    \end{pmatrix}
  %}^{(***)}
  \\
  \begin{pmatrix}
    c_{0} & 0 \\
    0 & c_{0}
  \end{pmatrix}
  \begin{pmatrix}
    \e^{-\I mc^2\tau/(2\hbar)} & 0 \\
    0 & \e^{+\I mc^2\tau/(2\hbar)}
  \end{pmatrix}
  % \begin{pmatrix}
  %   \Psi_1(\vec x, t) \\ \Psi_2(\vec x, t)
  % \end{pmatrix}
  +\bigO{\tau^3}
  \,.
\end{multline}

In a computer implementation of the real space split operator method,
the wave function $\Psi(\vec x, t)$ is discretized on a rectangular
\mbox{$(d+1)$-dimensional} space-time lattice with spatial grid spacings
$h_i$ ($i\in\{1,\dots,d\}$) and time steps of size $\tau$, viz.
\begin{subequations}
  \begin{align}
    \vec x_{\vec n} & =
    \begin{pmatrix}
      h_1n_1 \\ \vdots \\ h_d n_d
    \end{pmatrix}
    \quad\text{with $n_i \in\mathbb{Z}$,}\\
    t^k & = \tau k\quad\text{with $k\in\mathbb{Z}$.}
  \end{align}
\end{subequations}
The quantity $\vec h$ will refer to a \mbox{$d$-dimensional} vector with
components $h_i$. The discretized two-component wave function will be
denoted by
\begin{equation}
  \Psi(\vec x_{\vec n}, t^k) = 
  \begin{pmatrix}
     \Psi_1(\vec x_{\vec n}, t^k) \\
     \Psi_2(\vec x_{\vec n}, t^k)
   \end{pmatrix} =
  \begin{pmatrix}
     \Psi^k_{1, \vec n} \\
     \Psi^k_{2, \vec n}
   \end{pmatrix} =
   \Psi^k_{\vec n}
\end{equation}
and $\Psi^k$ identifies a vector with components $\Psi^k_{\vec n}$ for
fixed $k$.  The discrete version of the pseudo norm (\ref{eq:pseudo_norm})
reads
\begin{equation}
  \label{eq:pseudo_norm_d}
  \norm{\Psi^k}_{\sigma_3,\vec h}^2 =
  \prod_{i=1}^d h_i \sum_{\vec n} {\Psi^k_{\vec n}}^* \sigma_3 \Psi^k_{\vec n}\,.
\end{equation}
For the stability analysis we will also need the proper norm
\begin{equation}
  \label{eq:norm_d}
  \norm{\Psi^k}_{\vec h}^2 =
  \prod_{i=1}^d h_i \sum_{\vec n} {\Psi^k_{\vec n}}^* \Psi^k_{\vec n}\,.
\end{equation}

First and second order derivatives in (\ref{eq:op_U_split2})
are approximated by finite differences. We will utilize the second order
approximations $\nabla_{2,h} f(x)$ and $\nabla^2_{2,h} f(x)$
\begin{subequations}
  \label{eq:fdh2}
  \begin{align}
    \frac{\partial f(x)}{\partial x} & =
    \nabla_{2,h} f(x) + \bigO{h^2} =
    \frac{-f(x-h)+f(x+h)}{2h} + \bigO{h^2} \\
    \frac{\partial^2 f(x)}{\partial x^2} & =
    \nabla^2_{2,h} f(x) + \bigO{h^2} =
    \frac{f(x-h)-2f(x)+f(x+h)}{h^2} + \bigO{h^2} 
  \end{align}
\end{subequations}
as well as the fourth order schemes $\nabla_{4,h} f(x)$ and
$\nabla^2_{4,h} f(x)$
\begin{subequations}
  \label{eq:fdh4}
  \begin{align}
    \frac{\partial f(x)}{\partial x} & =
    \nabla_{4,h} f(x) + \bigO{h^4} =
    \frac{f(x-2h)-8f(x-h)+8f(x+h)-f(x+2h)}{12h} + \bigO{h^4} \\
    \frac{\partial^2 f(x)}{\partial x^2} & =
    \nabla^2_{4,h} f(x) + \bigO{h^4} =
    \frac{-f(x-2h)+16f(x-h)-30f(x)+16f(x+h)-f(x+2h)}{12h^2} + \bigO{h^4} \,.
  \end{align}
\end{subequations}
Plugging (\ref{eq:fdh2}) or (\ref{eq:fdh4}) into (\ref{eq:op_U_split2})
and taking into account boundary conditions yields the finite difference
scheme (the discretized version of (\ref{eq:U_split}))
\begin{equation}
  \label{eq:fds}
  \Psi^{k+1} = U_{\vec h, \tau} \Psi^k
\end{equation}
with some sparse $2N\times 2N$ matrix $U_{\vec h, \tau}$ that depends on
the grid spacings $\vec h$ and $\tau$, where $N$ denotes the total
number of spatial grid points. Thus, the real space split operator
method for the Klein-Gordon equation is essentially a matrix vector
multiplication. A stability analysis has to examine the properties of
this matrix vector multiplication as a function of $\vec h$ and
$\tau$. Before we can proceed with a stability analysis, however, we
have to give the term \emph{numerical stability} a precise definition.

\section{Strong and algebraic stability}
\label{sec:stability}

Loosely speaking numerical stability of a finite difference scheme means
that numerical errors that are caused by the discretization of a partial
differential equation are bounded. There is no unique way to cast this
intuitive notion into a formal mathematical statement and different
mathematical concepts of stability may be found in the literature. We
will consider the so-called strong stability
\cite{Lax:Richtmyer:1956:stability,Kreis:1962:Stabilitaet} and algebraic
stability \cite{Forsythe:Wasow:1960:FDM,Kreis:1962:Stabilitaet}.  An
explicit finite difference scheme $\Psi^{k+1} = U_{\vec h, \tau}\Psi^k$
for a partial differential equation, that is first-order in time, is
called strongly stable in the stability region $\Lambda$ if there is for
any fixed positive time $t$ a constant $C$ such that
\begin{equation}
  \norm{\Psi^k}_{\vec{h}}\leq C\norm{\Psi^0}_{\vec h}
  \qquad\text{for all}\quad 
  (\vec h,\tau)\in\Lambda\,,\quad 
  0\leq k\tau\leq t\,.
\end{equation}
Strong stability requires that the norm $\norm{\Psi^k}_{\vec{h}}$ is
bounded independently of the grid spacings $\tau$ and $\vec h$. An
explicit finite difference scheme $\Psi^{k+1} = U_{\vec h, \tau}\Psi^k$
for a first-order partial differential equation is called algebraically
stable in a stability region $\tilde\Lambda$ if there are for any fixed
positive time $t$ two constants $C$ and $\alpha\ge 0$ such that
\begin{equation}
  \norm{\Psi^k}_{\vec{h}}\leq C\tau^{-\alpha}\norm{\Psi^0}_{\vec h}
  \qquad\text{for all}\quad 
  (\vec h,\tau)\in\tilde\Lambda\,,\quad 
  0\leq k\tau\leq t\,.
\end{equation}
Algebraic stability allows the norm to grow proportionally to
$\tau^{-\alpha}$ and, therefore, it is not necessarily bounded in the
limit \mbox{$\tau\to0$.} Note that algebraic stability is a rather weak
stability concept; there is no intermediate regime between algebraic
stability and exponential explosion.  Strong stability implies algebraic
stability. For partial differential equations for scalar unknown
functions strong and algebraic stability are equivalent. For systems of
partial differential equations as the Klein-Gordon equation
(\ref{eq:Klein-Gordon-Schroedinger}), however, strong stability does not
follow by algebraic stability \cite{Kreis:1962:Stabilitaet}.

Strong and algebraic stability are closely related to convergence. A
finite difference scheme is called convergent if the discretization
error satisfies
\begin{equation}
  \norm{\Psi(\vec x, \tau k)-\Psi^k}_{\vec{h}}\rightarrow 0
  \qquad\text{for}\quad
  \vec h, \tau\rightarrow 0\,.
\end{equation}
Under certain conditions, strong and algebraic stability imply
convergence.  A consistent finite difference scheme for a partial
differential equation for which the initial value problem depends
continuously on the initial condition and that is of first order in its
time derivative is convergent if and only if it is strongly stable
\cite{Lax:Richtmyer:1956:stability}. An explicit finite difference
scheme $\Psi^{k+1}=U_{\vec h, \tau}\Psi^k$ for a first order partial
differential equation $\I\hbar{\partial \Psi(\vec x, t)}/{\partial
  t}=\op H(t)\Psi(\vec x, t)$ is called consistent if
\begin{equation}
  \norm{
    \frac{\Psi^{k+1}-\Psi^k}{\tau} - \frac{U_{\vec h, \tau}-1}{\tau}\Psi^k -
    \left(
      \frac{\partial \Psi(\vec x, t)}{\partial t} 
      -\frac{1}{\I\hbar}\op H\Psi(\vec x, t)
    \right)
  }_{\vec h}\to 0
  \qquad\text{for}\quad
  \vec h, \tau\rightarrow 0\,.
\end{equation}
Note that a consistent strongly stable scheme is convergent
independently of the initial condition.  If a finite difference scheme
$\Psi^{k+1}=U_{\vec h, \tau}\Psi^k$ is algebraically stable and $U_{\vec
  h, \tau}$ does not depend on spatial coordinates, then it is
convergent provided that the Fourier transform $\tilde\Psi(\vec p,0)$ of
the initial condition $\Psi(\vec x,0)$ decays faster than an appropriate
power $\abs{\vec p}^{-l}$ for $\abs{\vec p}\rightarrow\infty$
\cite{Kreis:1962:Stabilitaet}.  Note that this condition applies only to
homogeneous problems ($U_{\vec h, \tau}$ is not allowed to depend on
spatial coordinates.) and it is a sufficient condition for convergence
but not necessary. The condition on the initial function $\Psi(\vec x,
0)$ is satisfied by any function that is at least \mbox{$l$-times}
differentiable and its \mbox{$l$th} derivative is absolutely
integrable.

\section{Stability conditions}
\label{sec:Conditions}

There are various necessary and/or sufficient criteria
\cite{Lax:Richtmyer:1956:stability,Kreis:1962:Stabilitaet,Strikwerda:Wade:1997:Survey}
for strong or algebraic stability of finite difference schemes of the
type (\ref{eq:fds}). For example, one can show that a finite difference
scheme of the type (\ref{eq:fds}) is strongly stable in a region
$\Lambda$ if and only if for every time $t$, there exists a constant $C$
such that powers of $U_{\vec h, \tau}$ are bounded as
\begin{equation}
  \label{eq:strong_stability_cond}
  \norm{U_{\vec h, \tau}^k}\leq C
  \qquad\text{for all}\quad
  (\vec h,\tau)\in\Lambda,\quad 0\leq k\tau\leq t\,.
\end{equation}
For a stability analysis in Fourier space one phrases the discrete
function $\Psi^k$ via its Fourier transform $\tilde\Psi^k(\vec p)$, viz.
\begin{equation}
  \label{eq:Psi_Fourier}
  \Psi^k_{\vec n} =
  \frac{1}{(2\pi)^{d/2} \prod_{i=1}^d h_i}
  \int_{-\pi}^{\pi}\cdots\int_{-\pi}^{\pi}
  \exp\left(
    \I\sum_{i=1}^d n_i \xi_i
  \right)\tilde\Psi^k\left(\vec \xi \right)\,\D^d \xi \,.
\end{equation}
Plugging the ansatz (\ref{eq:Psi_Fourier}) into a finite difference
scheme of type (\ref{eq:fds}) for a partial differential equation with
constant coefficients and assuming a spatial grid of infinite size
individual Fourier modes decouple yielding a propagation equation
\begin{equation}
  \label{eq:fds_Fourier}
  \tilde\Psi^{k+1}(\vec\xi) = 
  \tilde U_{\vec h, \tau}(\vec\xi) \tilde\Psi^k(\vec\xi)
\end{equation}
for each Fourier mode $\vec\xi\in[-\pi,\pi]^d$.  The equivalent to the condition
(\ref{eq:strong_stability_cond}) in Fourier space reads
\begin{equation}
  \label{eq:strong_stability_cond_Fourier}
  \norm{\tilde U_{\vec h, \tau}(\vec\xi)^k}\leq \tilde C
  \qquad\text{for all}\quad
  \vec\xi\in[-\pi,\pi]^d,\quad
  (\vec h,\tau)\in\Lambda,\quad 0\leq k\tau\leq t\,.
\end{equation}
The condition (\ref{eq:strong_stability_cond_Fourier}) follows from 
(\ref{eq:strong_stability_cond}) via Parseval's Theorem.

Proving the boundedness of powers of a matrix is often not easy, thus
other criteria may be applied. The von-Neumann criterion requires for a
one-step finite difference scheme with constant coefficients
(\ref{eq:fds}) that there is a constant $K$ such that for all $(\vec
h,\tau)$ in the stability region $\tilde\Lambda$ the eigenvalues
$u_{\vec h, \tau}(\vec\xi)$ of $\tilde U_{\vec h, \tau}(\vec\xi)$ are
bounded as
\begin{equation}
  \label{eq:von_Neumann}
  \abs{u_{\vec h, \tau}(\vec\xi)}\leq1+K\tau \,.
\end{equation}
The von-Neumann stability criterion provides a sufficient condition for
algebraic stability \cite{Kreis:1962:Stabilitaet}. Because strong
stability implies algebraic stability, the von-Neumann stability
criterion is a necessary criterion for strong stability but in general
not sufficient.  In the stability analysis of Sec.~\ref{sec:Neumann} we
will use the slightly stronger criterion $\abs{u_{\vec h,
    \tau}(\vec\xi)}\leq1$.

% A sufficient condition for strong stability is fulfilled if the spectral
% norm $\norm{\tilde U_{\vec h, \tau}(\vec\xi)}$ of the matrix $\tilde
% U_{\vec h, \tau}(\vec\xi)$ is bounded by
% \begin{equation}
%   \label{eq:strong_sufficient}
%   \norm{\tilde U_{\vec h, \tau}(\vec\xi)} = 
%   \max_i \sqrt{\abs{\lambda_i}} \le 1 + C\tau 
%   \qquad\text{for all}\quad
%   \vec\xi\in[-\pi,\pi]^d\,,
% \end{equation}
% where $\lambda_i$ denote the eigen-values of $\tilde U_{\vec h,
%   \tau}(\vec\xi)^*\tilde U_{\vec h, \tau}(\vec\xi)$. This sufficient condition
% follows from the inequalities
% \begin{multline}
%   \norm{\tilde\Psi^{k}(\vec\xi)}_{\vec h} = 
%   \norm{\tilde U_{\vec h, \tau}(\vec\xi)^k \tilde\Psi^0(\vec\xi)}_{\vec h} \le
%   \norm{\tilde\Psi^{0}(\vec\xi)}_{\vec h}\norm{\tilde U_{\vec h, \tau}(\vec\xi)^k} \le
%   \norm{\tilde\Psi^{0}(\vec\xi)}_{\vec h}\norm{\tilde U_{\vec h, \tau}(\vec\xi)}^k \le \\
%   \norm{\tilde\Psi^{0}(\vec\xi)}_{\vec h}(1+C\tau)^k \le
%   \norm{\tilde\Psi^{0}(\vec\xi)}_{\vec h}\left(1+C\frac{t}{k}\right)^k \le
%   \norm{\tilde\Psi^{0}(\vec\xi)}_{\vec h}\e^{Ct}\,.
% \end{multline}
% If $\tilde U_{\vec h, \tau}(\vec\xi)$ and its Hermitian transpose
% $\tilde U_{\vec h, \tau}(\vec\xi)^*$ commute then absolute values of the
% eigen-values of $\tilde U_{\vec h, \tau}(\vec\xi)^*\tilde U_{\vec h,
%   \tau}(\vec\xi)$ are the squared values of absolute values of the
% eigen-values of $\tilde U_{\vec h, \tau}(\vec\xi)$. Thus, the
% von-Neumann stability condition is also sufficient for strong stability
% in this case.

\section{Von-Neumann stability analysis for homogeneous potentials}
\label{sec:Neumann}

In this section we are going to perform a von-Neumann stability analysis
for the real space split operator method for the Klein-Gordon equation
with homogeneous potentials
\begin{equation}
  \label{eq:const_pot}
  \phi(\vec x, t) = \phi_0 \qquad
  \vec A(\vec x, t) = \vec A_0 \,.
\end{equation}
Transforming the matrix (\ref{eq:op_U_split2}) with finite differences
of second order (\ref{eq:fdh2}) or fourth order (\ref{eq:fdh4}) into
Fourier space the finite difference operators in (\ref{eq:op_U_split2})
have to be replaced by
\begin{subequations}
  \label{eq:fdh2_Fourier}
  \begin{align}
    \nabla_{3,\vec h} f(\vec x) & \to
    \tilde\nabla_{3,\vec h} f(\vec\xi) =
    -\I
    \begin{pmatrix}
    \frac{\sin(\xi_1)}{h_1} \\
    \vdots\\
    \frac{\sin(\xi_d)}{h_d} 
  \end{pmatrix}f(\vec\xi)\\
    \nabla^2_{3,\vec h} f(\vec x) & \to
    \tilde\nabla^2_{3,\vec h} f(\vec\xi) =
    -\sum_{i=1}^d \frac{2-2\cos(\xi_i)}{h_i^2}f(\vec\xi) 
  \end{align}
\end{subequations}
or
\begin{subequations}
  \label{eq:fdh4_Fourier}
  \begin{align}
    \nabla_{5,\vec h} f(\vec x) & \to
    \tilde\nabla_{5,\vec h} f(\vec\xi) =
    -\I
    \begin{pmatrix}
    \frac{8\sin(\xi_1)-\sin(2\xi_1)}{6h_1} \\
    \vdots\\
    \frac{8\sin(\xi_d)-\sin(2\xi_d)}{6h_d} \\
  \end{pmatrix}f(\vec\xi)\\
    \nabla^2_{5,\vec h} f(\vec x) & \to
    \tilde\nabla^2_{5,\vec h} f(\vec\xi) =
    -\sum_{i=1}^d \frac{15-16\cos(\xi_i)+\cos(2\xi_i)}{6h_i^2}f_i(\vec\xi) \,,
  \end{align}
\end{subequations}
respectively. With (\ref{eq:fdh2_Fourier}) and (\ref{eq:fdh4_Fourier})
the Fourier space propagation matrix becomes for homogeneous potentials 
\begin{equation}
  \label{eq:U_split_F_const}
  \tilde U_{\vec h, \tau}(\vec\xi) =
  \begin{pmatrix}
    \e^{-\I mc^2\tau/\hbar}\left(1+\left(
        \frac{\I\tau\hbar}{2m} \tilde\nabla_{n, \vec h}^2 + 
        \frac{\tau q\vec A_0}{m} \cdot \tilde\nabla_{n, \vec h} - 
        \frac{\I\tau q^2\vec A_0^2}{2m\hbar}
      \right)\right) &
    \phantom{0}+\left(
      \frac{\I\tau\hbar}{2m} \tilde\nabla_{n, \vec h}^2 + 
      \frac{\tau q\vec A_0}{m} \cdot \tilde\nabla_{n, \vec h} -
      \frac{\I\tau q^2\vec A_0^2}{2m\hbar}
    \right) \\[2ex]
    \phantom{0}-\left(
      \frac{\I\tau\hbar}{2m} \tilde\nabla_{n, \vec h}^2 + 
      \frac{\tau q\vec A_0}{m} \cdot \tilde\nabla_{n, \vec h} - 
      \frac{\I\tau q^2\vec A_0^2}{2m\hbar}
    \right) &
    \e^{\I mc^2\tau/\hbar}\left(1-\left(
        \frac{\I\tau\hbar}{2m} \tilde\nabla_{n, \vec h}^2 + 
        \frac{\tau q\vec A_0}{m} \cdot \tilde\nabla_{n, \vec h} -
        \frac{\I\tau q^2\vec A_0^2}{2m\hbar}
      \right)\right)
  \end{pmatrix}
  \e^{-\I q\phi_0\tau/\hbar}\,,
\end{equation}
where $n\in\{2,4\}$. To derive (\ref{eq:U_split_F_const}) we have used
that for a homogeneous scalar potential the matrix $(*)$ in
(\ref{eq:op_U_split2}) commutes with all other matrices in
(\ref{eq:op_U_split2}).  The moduli of the eigenvalues of
(\ref{eq:U_split_F_const}) do not depend on the factor
$\e^{-\I q\phi_0\tau/\hbar}$. Therefore, for a von-Neumann stability
analysis we can set $\phi_0=0$ without loss of generality. The numerical
stability does not depend on the magnitude of the scalar potential. For
convenience, we introduce the real valued quantities
\begin{align}
  \kappa(\vec\xi) & = \sum_{i=1}^d \kappa_i(\xi_i) 
  \qquad\text{with}\qquad \kappa_i(\xi_i) =
  \frac{\hbar}{2m}\tilde\nabla_{n,h_i}^2 + 
  \frac{q A_{0,i}}{\I m}\tilde\nabla_{n,h_i} -
  \frac{q^2 A_{0,i}^2}{2m\hbar} \\
  \intertext{and}
  \gamma(\vec\xi) & =
  \cos\left(mc^2\tau/\hbar\right) +
  \kappa(\vec\xi)\tau\sin\left(mc^2\tau/\hbar\right)
\end{align}
such that the eigen-values of (\ref{eq:U_split_F_const}) are given by
\begin{equation}
  u_{\vec h, \tau}(\vec\xi) =
  \gamma(\vec\xi) \pm \sqrt{\gamma(\vec\xi)^2 -1 } \,.
\end{equation}
The quantity $\gamma(\vec\xi)$ is real valued and, therefore, the
von-Neumann criterion $\abs{u_{\vec h, \tau}(\vec\xi)}\leq1$ is
equivalent to $\gamma(\vec\xi)^2\leq 1$ which must be fulfilled for
every $\vec\xi\in[-\pi,\pi]^d$.  The von-Neumann criterion is fulfilled
for all $\vec\xi\in[-\pi,\pi]^d$ if it is fulfilled for all $\vec\xi^*$
that make $\gamma(\vec\xi)^2$ extremal. Qualification for extremality of
$\gamma(\vec\xi)^2$ is
\begin{equation}
  \left.
    \frac{\partial\left( \gamma(\vec\xi)^2\right)}{\partial \xi_i} 
  \right|_{\vec\xi=\vec\xi^*} = 
  % 2\left(
  %   \cos\left(mc^2\tau/\hbar\right) +
  %   \kappa(\vec\xi^*)\tau\sin\left(mc^2\tau/\hbar\right)
  % \right)\tau\sin\left(mc^2\tau/\hbar\right) 
  % \frac{\D\kappa_i(\xi_i^*)}{\D \xi_i^*} = 0
  2\gamma(\vec\xi^*)\tau\sin\left(mc^2\tau/\hbar\right) 
  \left.
    \frac{\D\kappa_i(\xi_i)}{\D \xi_i} 
  \right|_{\xi_i=\xi^*_i} = 0
  \qquad\text{for all $i\in\{1,\dots,d\}$}\,.
\end{equation}
Thus, to maximize $\gamma(\vec\xi)^2$
\begin{equation}
  \left.
    \frac{\D\kappa_i(\xi_i)}{\D \xi_i} 
  \right|_{\xi_i=\xi^*_i} = 0
  \qquad\text{for all $i\in\{1,\dots,d\}$}
\end{equation}
is required. The region $\tilde\Lambda$ of algebraic stability for the
scheme (\ref{eq:fds}) with constant potentials is finally given by
\begin{equation}
  \label{eq:Lambda}
  \tilde\Lambda = \left\{
    (\vec h,\tau)\in [0,\infty)^{d+1}
    \quad\text{%
      with $\gamma(\vec\xi^*)^2\leq1$, 
      for all $\vec\xi^*\in[-\pi,\pi]^d$ that are the solutions of $\left.
        \frac{\D\kappa_i(\xi_i)}{\D \xi_i} 
      \right|_{\xi_i=\xi^*_i}=0$} 
  \right\}\,.
\end{equation}

\begin{figure}
  \centering
  \includegraphics[scale=0.45]{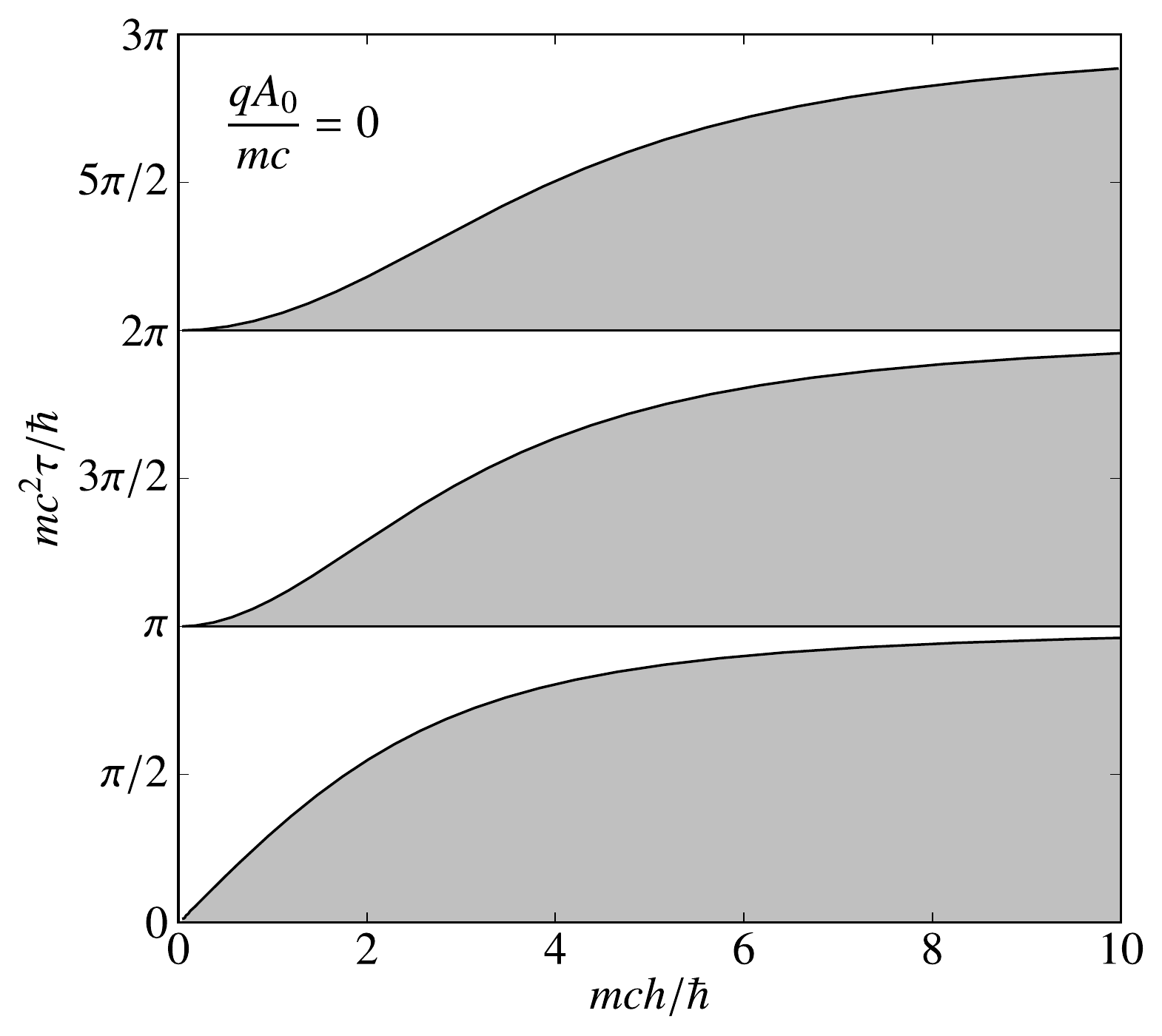}\qquad%
  \includegraphics[scale=0.45]{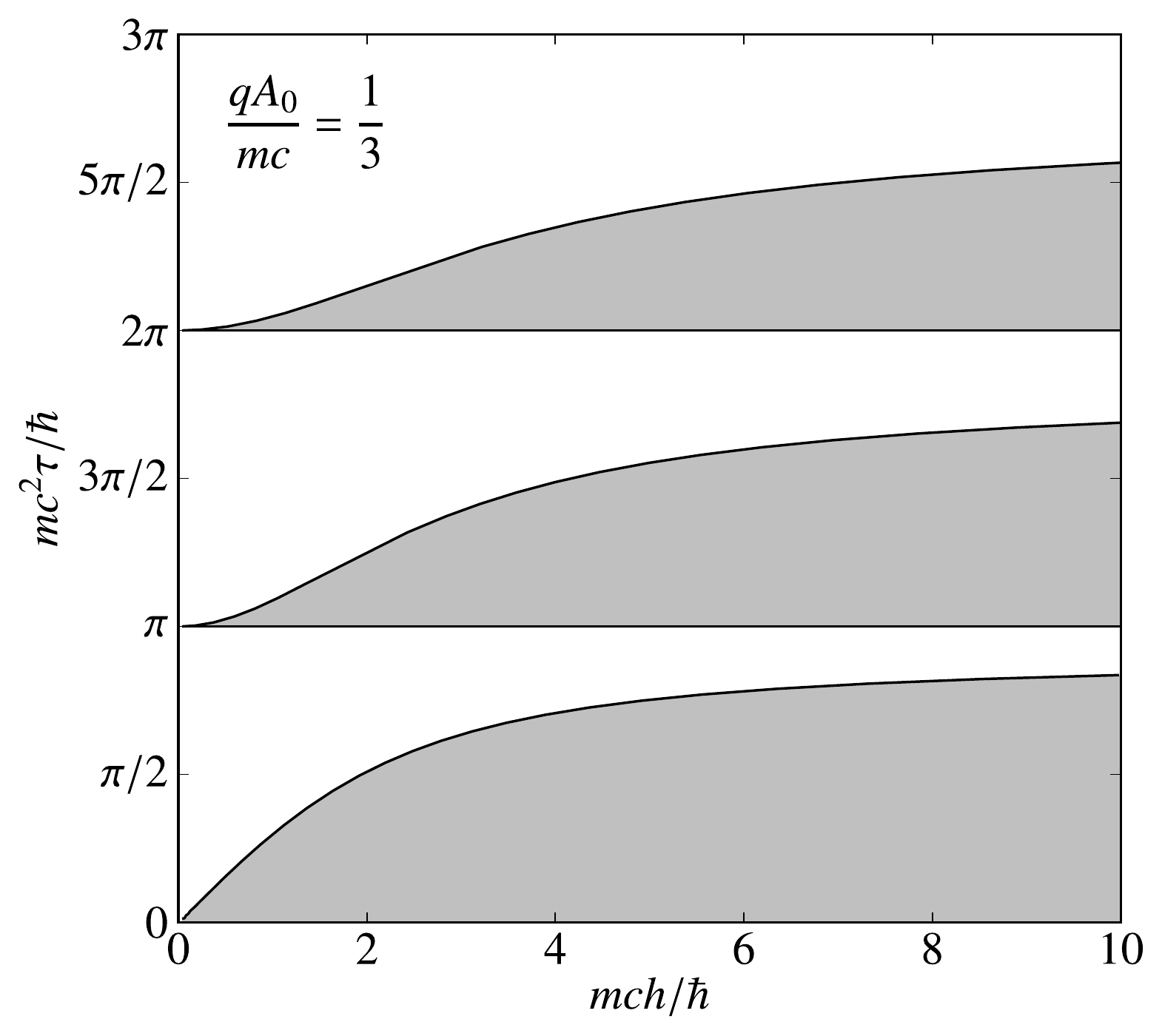}
  \caption{Regions of algebraic stability $\tilde\Lambda$ (shaded in
    gray) for a one-dimensional finite difference scheme with three
    point finite differences (\ref{eq:fdh2}) for two different values of
    the homogeneous vector potential $A_0$ as a function of the spatial
    grid spacing $h$ and the temporal step width $\tau$. The larger the
    vector potential the smaller the stability region.}
  \label{fig:Lambda_1d}
\end{figure}

Figure~\ref{fig:Lambda_1d} shows the region of algebraic stability
$\tilde\Lambda$ for a one-dimensional finite difference scheme with
three point finite differences (\ref{eq:fdh2}) for two different values
of the homogeneous vector potential as a function of the spatial grid
spacing $h$ and the temporal step width $\tau$. The stability region
separates into an infinite number of disconnected domains. From a
practical point of view, only the stability domain that includes the
point $(h, \tau)=(0, 0)$ is relevant because the temporal step width
$\tau$ is not only restricted by the stability criterion
(\ref{eq:Lambda}) but also by $mc^2\tau/\hbar<\pi$. The later criterion
follows from the wave-function's fast phase oscillations that are
proportional to the typical energy of a wave packet which is about
$mc^2$. In stability domains with $mc^2\tau/\hbar>\pi$ the finite
difference scheme is stable but does not give accurate numerical
solutions.  The shape of the stability domains depends on the modulus of
the vector potential. The larger the vector potential the smaller the
stability domains. The region of algebraic stability for a finite
difference scheme based on five point formulas (\ref{eq:fdh4}) looks
qualitatively similar to the region for three point formulas
(\ref{eq:fdh2}) but the stability domains are slightly smaller, see
Fig.~\ref{fig:Lambda_1d_num}.

\begin{figure}
  \centering
  \includegraphics[scale=0.45]{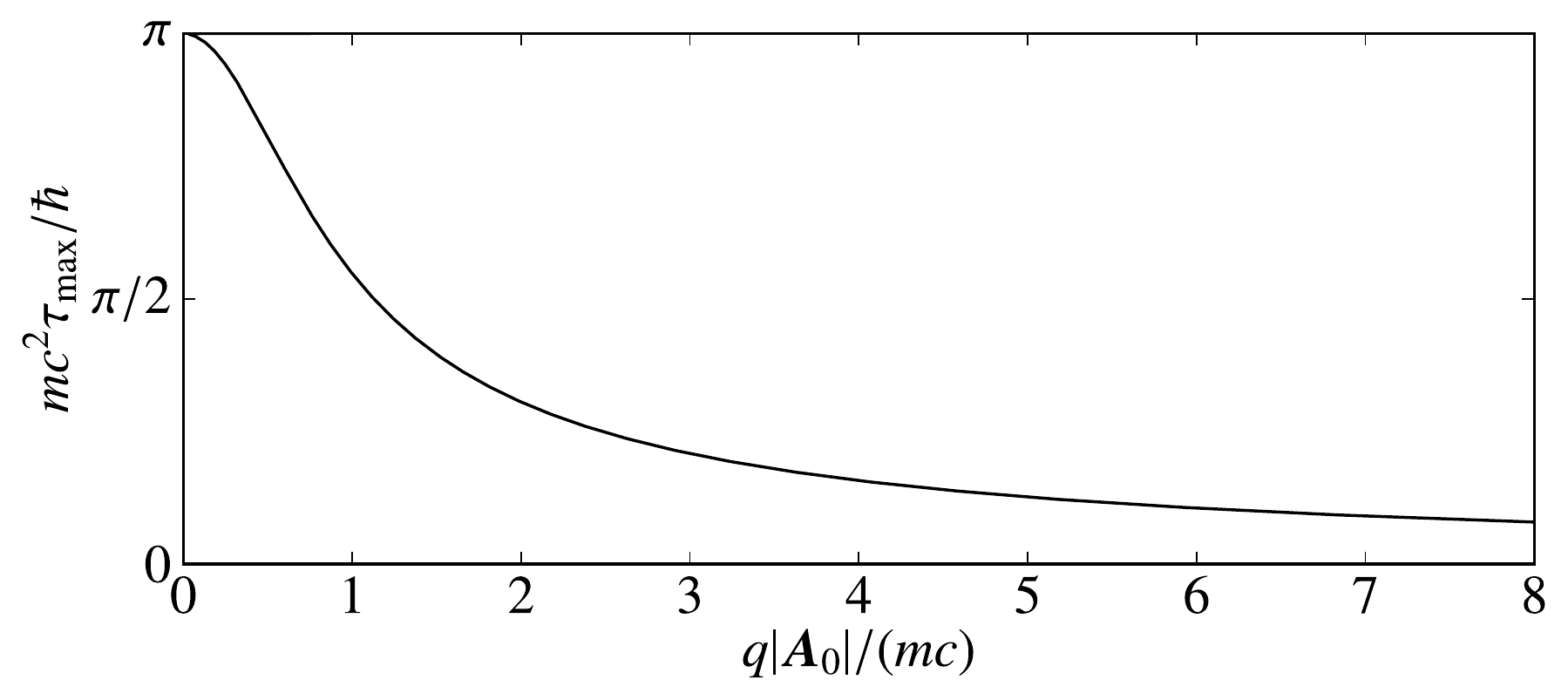}%
  \caption{Upper bound $\tau_\mathrm{max}$ on the temporal step width
    $\tau$ as a function of the magnitude of the vector potential $\vec
    A_0$.}
  \label{fig:tau_max}
\end{figure}

It is instructive to analyze the region of algebraic stability
$\tilde\Lambda$ in some limiting cases. Let us consider the case of
small step widths and small vector potentials with
\begin{equation*}
  \frac{mc^2\tau}{\hbar}\ll 1\,,\qquad
  \frac{mch_i}{\hbar}\ll 1\,,\qquad
  \frac{q|A_{0,i}|}{mc}\ll 1\qquad
  \text{for all $i\in\{1,\dots,d\}$}
\end{equation*}
first. For second order finite differences (\ref{eq:fdh2}) the condition
$\gamma(\vec\xi^*)^2\leq 1$ yields in this limit in leading order
\begin{equation}
  \left(
    1 -
    \frac{mc^2\tau}{\hbar}\frac{1}{2}
    \sum_{i=1}^d\left(\frac{mch_i}{\hbar}\right)^{-2}(2-2\cos(\xi_i))
    \frac{mc^2\tau}{\hbar}  
  \right)^2\leq 1\,.
\end{equation}
This expression becomes maximal for $\xi_i^*=\pm\pi$. For equal grid
spacings $h_1=\ldots=h_d=h$ the condition $\gamma(\vec\xi^*)^2\leq1$
simplifies to
\begin{equation}
  \left(1 - 2d\left(\frac{c\tau}{h}\right)^2 \right)^2 \leq 1
\end{equation}
which is equivalent to 
\begin{equation}
  \frac{h}{\tau}>\sqrt{d}c\,.
\end{equation}
A similar calculation for five point finite differences formulas
(\ref{eq:fdh4}) yields the condition
\begin{equation}
  \label{eq:Lambda_zero_fields_fdh4}
  \frac{\sqrt{3}}{2}\frac{h}{\tau}>\sqrt{d}c\,.
\end{equation}
The other limit we want to consider is the limit of large spatial step
sizes with 
\begin{equation*}
  \frac{mch_i}{\hbar}\gg 1
  \,.
\end{equation*}
In this limit the condition $\gamma(\vec\xi^*)^2\leq 1$ yields for
second order finite differences (\ref{eq:fdh2}) as well as for fourth
order finite differences (\ref{eq:fdh4})  the inequality 
\begin{equation}
  \label{eq:tau_upper}
  \left(
    \cos\frac{mc^2\tau}{\hbar} - 
    \frac{1}{2}\frac{mc^2\tau}{\hbar}
    \left(\frac{q\vec A_0}{mc}\right)^2
    \sin\frac{mc^2\tau}{\hbar} 
  \right)^2\le1
\end{equation}
which no longer depends on $\vec\xi^*$.  The larger the spatial steps
the larger the allowed temporal step width. Thus (\ref{eq:tau_upper})
implies an upper bound on the temporal step width which we denote by
$\tau_\mathrm{max}$. Figure~\ref{fig:tau_max} shows $\tau_\mathrm{max}$
as a function of the magnitude of the vector potential $\vec A_0$. For a
vanishing vector potential we have $mc^2\tau_\mathrm{max}/\hbar=\pi$ and
$mc^2\tau_\mathrm{max}/\hbar<\pi$ otherwise.

\section{Numerical results}
\label{sec:numerics}

A von-Neumann stability analysis does not account for boundary
conditions, it implicitly assumes an infinite grid. Furthermore, the
von-Neumann stability analysis is not applicable to systems with
spatially dependent potentials. Thus, we are going to compare our
analytical results of Sec.~\ref{sec:Neumann} with the stability of
actual numerical simulations that work on finite grids and implement the
boundary condition that the wave function vanishes at border.  We will
determine the numerical stability for systems with spatially dependent
potentials as a function of the size of the grid spacings and compare
the results with the predictions of our von-Neumann stability analysis
for homogeneous potentials. In all our numerical simulations atomic
units (a.u.)  with $\hbar=1\,$a.u., $q=e=1\,$a.u., $m=1\,$a.u., and
$c\approx137.036\,$a.u.\ will be applied.

\subsection{Systems with homogeneous potentials}

\begin{figure}
  \centering
  \includegraphics[scale=0.45]{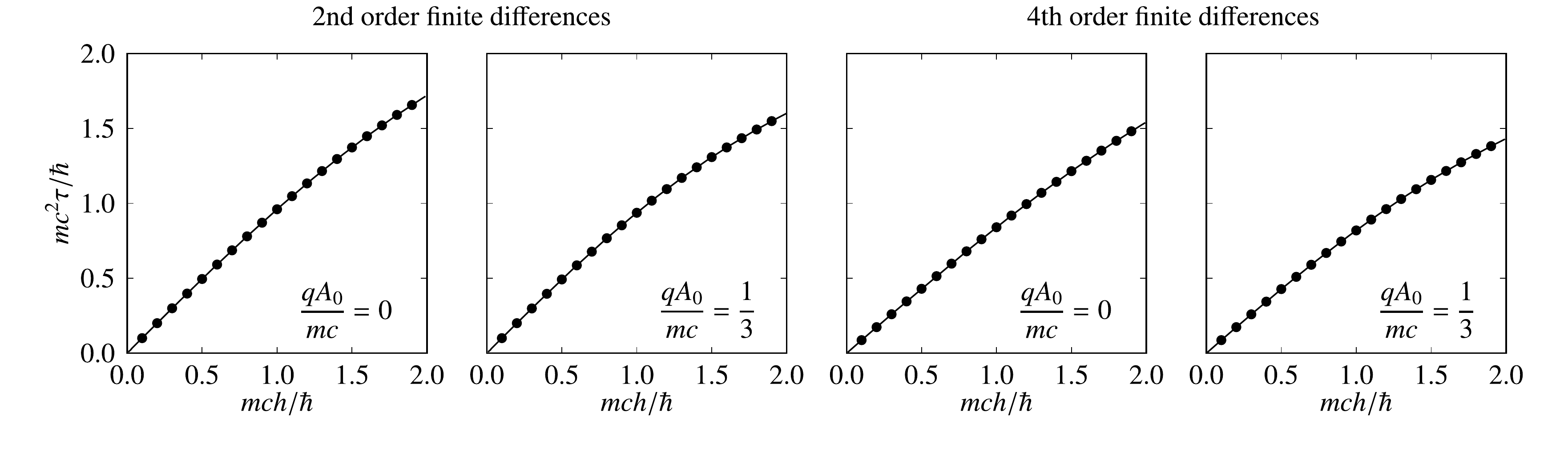}
  \caption{Comparison of the critical lines as predicted by
    (\ref{eq:Lambda}) (solid lines) with the points of transition from
    stability to instability as they have been observed in actual
    numerical simulations (filled circles) of one-dimensional wave
    packets using finite differences of second order (\ref{eq:fdh2}) and
    of fourth order (\ref{eq:fdh4}), respectively.}
  \label{fig:Lambda_1d_num}
\end{figure}

In order to check the condition~(\ref{eq:Lambda}) we determined the
region of stability also numerically by propagating a one-dimensional
wave packet in homogeneous potentials using different grid spacings $h$
and $\tau$.  The initial wave packet was given by a Gaussian
superposition of positive energy free particle eigen-states with
momentum $p$
\begin{equation}
  \Psi(x, 0) = \frac{1}{\sqrt{2\pi\hbar}}
  \int_{-\infty}^{\infty} 
  \frac{1}{2({1+p^2/(mc)^2})^{1/4}}
  \begin{pmatrix}
    1+\sqrt{1+p^2/(mc)^2} \\ 1-\sqrt{1+p^2/(mc)^2} 
  \end{pmatrix}
  \frac{1}{(2\pi\Delta^2)^{1/4}} 
  \exp\left(-\frac{(p-\bar p)^2}{4\Delta^2} + \I\frac{xp}{\hbar}\right) 
  \,\D p \,.
\end{equation}
The initial wave packet has mean momentum $\bar p$ and momentum width
$\Delta$. For our tests we chose $\bar p=20\,$a.u., $\Delta=1\,$a.u. and
a computational grid that ranges from $x_\mathrm{left}=-3\,$a.u.\ to
$x_\mathrm{right}=5\,$a.u. The wave packet was propagated from time
$t_\mathrm{init}=0\,$a.u.\ to $t_\mathrm{end}=1/20\,$a.u. For a free wave
packet not only the pseudo norm~(\ref{eq:pseudo_norm}) is conserved but
also the Euclidean norm~(\ref{eq:norm}). Thus, numerical stability may
be checked by determining if the discrete norm~(\ref{eq:norm_d}) grows
exponentially or not as a function of the grid spacings $h$ and $\tau$.
Figure~\ref{fig:Lambda_1d_num} presents our results for two different
vector potentials and finite difference formulas of second order
(\ref{eq:fdh2}) and of fourth order (\ref{eq:fdh4}), respectively. We
observe an excellent correspondence between the von-Neumann stability
analysis result (\ref{eq:Lambda}) and our numerical findings.

\subsection{Systems with spatially dependent potentials}

We considered so far systems with homogeneous potentials only. In
numerical calculations, however, one typically wants to simulate systems
with spatially dependent potentials to which the technique of the
von-Neumann stability analysis is not applicable. It is possible to show
that the results of von-Neumann stability analysis for systems with
homogeneous potentials can generally not be extended to inhomogeneous
systems. In particular, one can devise finite difference schemes for
partial differential equations that are instable although the
von-Neumann condition (\ref{eq:von_Neumann}) is satisfied locally
everywhere \cite{Kreis:1962:Stabilitaet}.

Assuming that the stability properties of the real space split operator
method are well behaved one expects that for systems with potentials
that vary on length scales that are large compared to the size of the
wave packet the stability region is given in close approximation by
(\ref{eq:Lambda}). If the stability is actually affected by the
potentials' spatial dependence then it is expected to have a larger
impact, if the potentials vary on the length scale that is given by the
wave-function's width. In order to test how a spatially dependent vector
potential changes the stability properties of the real space split
operator method we simulated scattering of a free wave packet on a
strong laser pulse having a very short wave length.

\begin{figure}
  \centering
  \parbox[c]{0.4\textwidth}{%
    \includegraphics[scale=0.45]{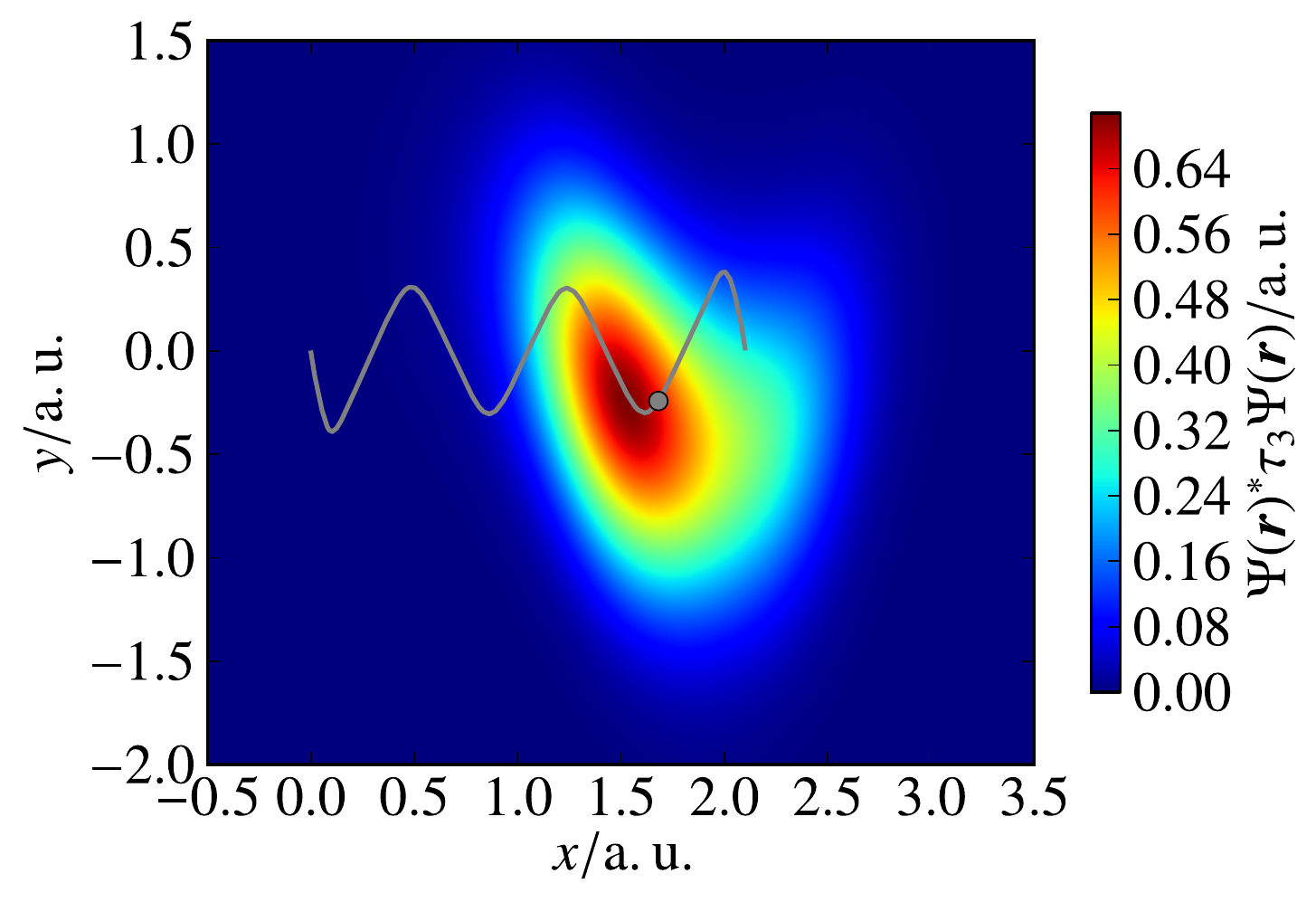}%
  }%
  \qquad\qquad%
  \parbox[c]{0.4\textwidth}{%
    \includegraphics[scale=0.45]{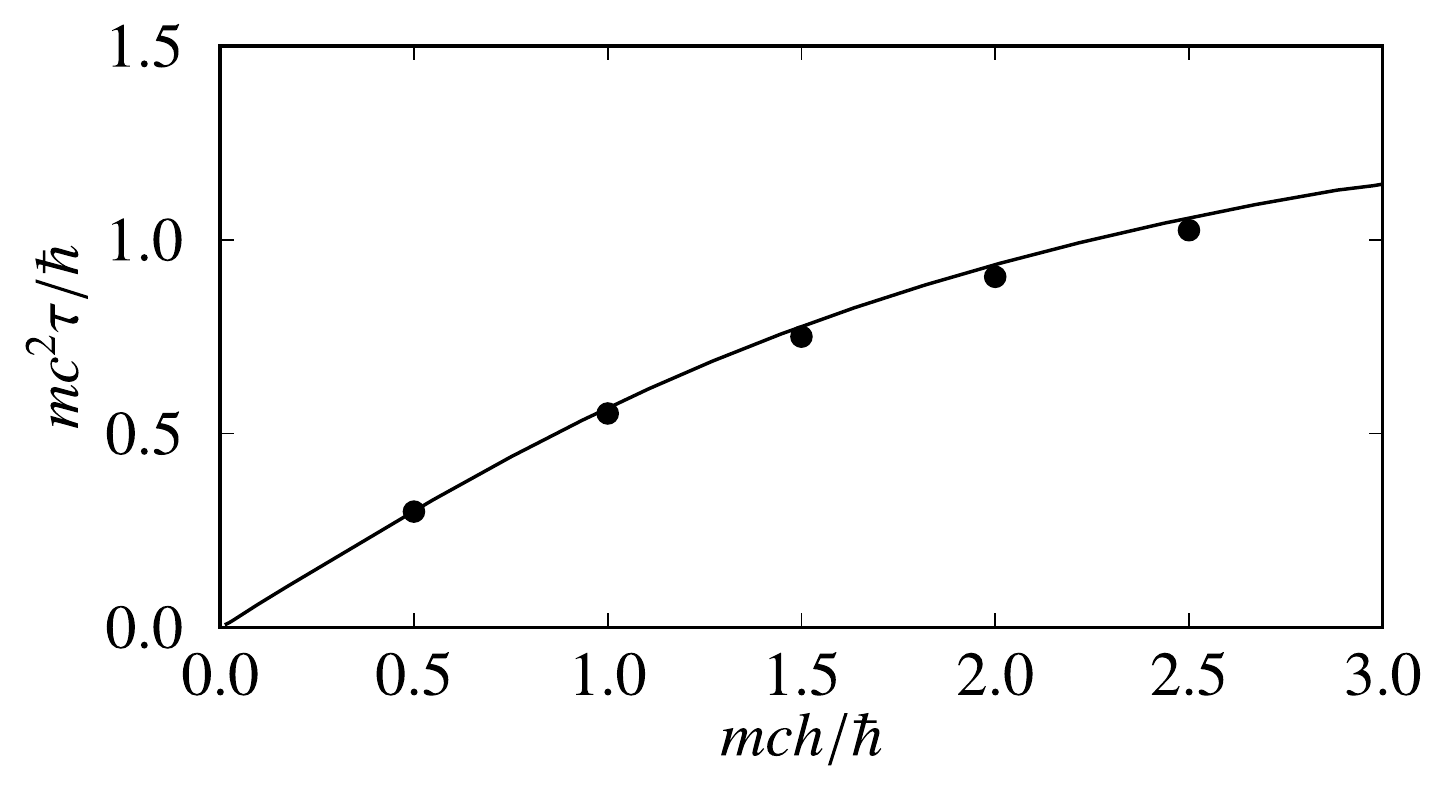}%
  }%
  \caption{Left panel: Motion of a wave packet in a ultra strong laser
    field. Plot shows the center of mass motion (gray line) and a
    snapshot of the density (\ref{eq:kg_dens}) with the corresponding
    center of mass (gray circle). Right panel: Stability line for the
    system in the left panel using fourth order finite differences. Dots
    indicate critical points as determined numerically, the solid line
    is given by (\ref{eq:Lambda}) using the vector-potential's maximal
    value.}
  \label{fig:Lambda_2d_laser_num}
\end{figure}

We start with a two-dimensional Gaussian wave packet
\begin{equation}
  \Psi(\vec x, 0) = \frac{1}{{2\pi\hbar}}
  \int_{-\infty}^{\infty} 
  \int_{-\infty}^{\infty} 
  \frac{1}{2({1+\vec p^2/(mc)^2})^{1/4}}
  \begin{pmatrix}
    1+\sqrt{1+\vec p^2/(mc)^2} \\ 1-\sqrt{1+\vec p^2/(mc)^2} 
  \end{pmatrix}
  \frac{1}{(2\pi\Delta^2)^{1/2}} 
  \exp\left(-\frac{\vec p^2}{4\Delta^2} + 
    \I\frac{\vec x\vec p}{\hbar}\right) 
  \,\D \vec p \,.
\end{equation}
The wave packet evolves in a laser pulse of wave length $\lambda$ with
the electromagnetic fields
\begin{subequations}
  \label{eq:laserpulse}
  \begin{align}
    \vec E(\vec x, t) & = 
    \vec E_0 \sin(\vec k\cdot\vec x - \omega t)
    f_{j, l}(\vec k\cdot\vec x - \omega t)\,, \\
    \vec B(\vec x, t) & = 
    \frac{\vec E_0}{c} \sin(\vec k\cdot\vec x - \omega t) 
    f_{j, l}(\vec k\cdot\vec x - \omega t)
    \intertext{with $|\vec k|=2\pi/\lambda=\omega/c$ and $\vec
      E_0\perp\vec k$ and an envelope function $f_{j, l}(\eta)$ of total
      $j$ half cycles including a linear turn-on ramp and a linear
      turn-off ramp of $l$ half cycles and a constant plateau in
      between, viz.}
    f_{j, l}(\eta) & = \begin{cases}
      (\eta+j\pi)/(l\pi) & \text{if $-j\le\eta/\pi\le-j+l$,} \\
      1 & \text{if $-j+l\le\eta/\pi\le-l$,} \\
      -\eta/(l\pi) & \text{if $-l\le\eta/\pi\le0$,} \\
      0 & \text{else.}
    \end{cases}
  \end{align}
\end{subequations}
The corresponding vector potential follows from (\ref{eq:laserpulse})
via
\begin{equation}
  \label{eq:A_laser}
  \vec A(\vec x, t) = - \int_0^t \vec E(\vec x, t')\,\D t'\,.
\end{equation}
For our simulation we chose the parameters $\Delta=1\,\mathrm{a.\,u.}$,
$\lambda=3\,\mathrm{a.\,u.}$, $j=8$, $i=2$ and $|\vec A_0|=|\vec
E_0|/\omega=mc/q$. The initial wave packet has a width of a half Bohr
radius and scatters at an ultra strong laser pulse having wave length
only six times larger and a peak intensity of
$1.09\times10^{26}\,\mathrm{W/cm^2}$. The left panel of
Fig.~\ref{fig:Lambda_2d_laser_num} shows the center of mass motion and a
snapshot of the density (\ref{eq:kg_dens}).

Determining the numerical stability by propagating the wave function
using different step widths $h_1=h_2=h$ and~$\tau$ we find that the
stability region for this system with spatially dependent vector
potential agrees well with (\ref{eq:Lambda}) if in (\ref{eq:Lambda}) the
maximal value of (\ref{eq:A_laser}) is used, see right panel of
Fig.~\ref{fig:Lambda_2d_laser_num}. Numerical instability was detected
by determining if (\ref{eq:norm_d}) grows exponentially or not. Note
that for a wave packet in a laser field the norm (\ref{eq:norm_d}) is
not a conserved quantity. Our numerical findings suggest the conjecture
that the real space split operator method is stable also for
inhomogeneous potentials if the von-Neumann condition
(\ref{eq:von_Neumann}) is satisfied locally everywhere.

\section{Conclusions}

We analyzed the stability of a real space split operator method for the
time-dependent Klein-Gordon equation and found that the method is
conditionally stable. For homogeneous potentials the region of algebraic
stability was determined as (\ref{eq:Lambda}) via a von-Neumann
stability analysis.  Our numerical simulations support the conjecture
that the condition (\ref{eq:Lambda}) is also applicable for spatially
dependent vector potentials. In the limit of small spatial grid spacings
$h$ in all spatial directions and small temporal steps $\tau$ the
stability condition becomes $h/\tau>\sqrt{d}c$ for second order finite
differences and $\sqrt{3}h/(2\tau)>\sqrt{d}c$ for fourth order finite
differences, respectively. Two conditions that might be more handy for
practical applications. Our results may help to circumvent numerical
instabilities in the application of the real space split operator method
for the Klein-Gordon equation by choosing appropriate spatial grid
spacings and an appropriate temporal step width.

\section*{Acknowledgment}

We gratefully acknowledge stimulating discussions with Matthias Ruf.

\secreferences
\bibliography{stability}

\end{document}